\documentclass[a4paper,11pt]{article}
\usepackage{jheppub} % for details on the use of the package, please
\usepackage{graphicx, graphics, color}
\usepackage[T1]{fontenc} % if needed

\usepackage{amsmath, amssymb, bm, graphicx, graphics, color, mathrsfs}
\usepackage{epsfig}
\usepackage{dcolumn}% Align table columns on decimal point
\usepackage{bm}% bold math
\usepackage{tikz}
\usepackage{graphicx, graphics, color}
\usepackage{dcolumn}% Align table columns on decimal point
\usepackage{bm}% bold math
\usepackage{hyperref}% add hypertext capabilities
\usepackage[mathlines]{lineno}% Enable numbering of text and display math

\newcommand{\be}{\begin{equation}}
\newcommand{\ee}{\end{equation}}
\newcommand{\ben}{\begin{eqnarray}}
\newcommand{\een}{\end{eqnarray}}

\newcommand{\la}{{\lambda}}

\newcommand{\cO}{{\cal O}}

\newcommand{\cL}{{\cal L}}
\newcommand{\cH}{{\cal H}}

\newcommand{\p}{\partial}
\newcommand{\na}{\nabla}

\newcommand{\Lie}{{\cal L}}

\newcommand{\tA}{\tilde A}

\newcommand{\tG}{\tilde G}

\newcommand{\ga}{\gamma}

\newcommand{\tB}{{\tilde B}}
\newcommand{\talpha}{{\tilde \alpha}}

\title{\boldmath 
DC conductivities and Stokes flows in Dirac semimetals}

\author[1]{Marek Rogatko\note{rogat@kft.umcs.lublin.pl, marek.rogatko@poczta.umcs.lublin.pl}}
\affiliation{Institute of Physics \\
Maria Curie-Sk{\l}odowska University \\
20-031 Lublin, pl. Marii Curie-Sk{\l}odowskiej 1, Poland}
\abstract{In the holographic model of Dirac semimetals being the Einstein-Maxwell scalar gravity with the auxiliary $U(1)$-gauge field,
coupled to the ordinary Maxwell one by {\it kinetic mixing } term, the black brane response to the electric fields and temperature gradient
has been elaborated. Using the foliation by hypersurfaces of constant radial coordinate we derive the exact form of the Hamiltonian and equations of motion in
the considered phase space. Examination of the Hamiltonian constraints enables us, to the leading order expansion of the linearised perturbations 
at the black brane event horizon, to derive Stokes equations for incompressible doubly charged fluid. Solving the aforementioned equations, one arrives at
the DC conductivities for the holographic Dirac semimetals.
  }

\keywords{Gauge-gravity correspondence,
Holography and condensed matter physics (AdS/CMT), Black Holes}

\begin{document} 

\maketitle
\flushbottom

%%%%%%%%%%%%%%%%%%%%%%%%%%%%%%%%%%%%%%%%%%%%%%%%%%%%%%%%%%%%%%%%%%%%%%%%%%%

%%%%%%%%%%%%%%%%%%%%%%%%%%%%%%%%%%%%%%%%%%%%%%%%%%%%%%%%%%%%%%%%%%%%%%%%%%%

\section{Introduction}
\label{sec:intro}
There has been observed a major resurgence of theoretical interests in connections among theories of gravity and Navier-Stokes equations governing
fluid dynamics, being the development of the idea of the so-called black hole membrane paradigm \cite{pri86}. Approximate solutions of the gravity theory 
relations were achieved by solving relativistic hydrodynamics equations, by the derivative expansion. Moreover the Navier-Stokes equations
begun to attract more attention in strongly correlated systems, via AdS/CFT correspondence \cite{zaanen-book}-\cite{nas15}. In \cite{bha09} the equations of relativistic hydrodynamics were reduced to the
incompressible Navier-Stokes equation in a particular scaling limit. Implementation of this scaling limit to holographically
induced fluid dynamics enabled to find gravity dual descriptions of an arbitrary solution of the forced non-relativistic incompressible Navier-Stokes equation.
In \cite{fou08} the hydrodynamics of relativistic conformal fluid at finite temperature was studied. Among all it was revealed that for viscous hydrodynamics in the limit of 
slow motion, equations could be cast to non-relativistic, incompressible Navier-Stokes ones. On the other hand, it was shown \cite{eli09} that such kind of relations 
could be derived from black hole membrane dynamics. 

In \cite{bre12} it was shown how the solutions of Navier-Stokes equations on hypersurface in Minkowski spacetime lead to the solutions of Einstein theory of gravity.

In the near-horizon and non-relativistic limits it was found that the perturbations effects in massive 
gravity in the bulk could be governed by incompressible Navier-Stokes equation \cite{pan16}. The implications of the certain modes to the vacuum solutions being consistent with
the hydrodynamic scaling, was elaborated in \cite{de19}. It happens that the inclusion in question corresponds to the solutions with certain types of matter. This fact reveals that 
gravity has description not only on null surfaces but also on timelike ones. The new setting of Navier-Stokes equations problem was delivered in
\cite{de19a}, where a metric constructed with a help of the scaling and symmetry properties Navier-Stokes relations was proposed.

Recently it has been also revealed in Einstein-Maxwell scalar gravity with a potential, holographically dual to the conformal field theory, in an asymptotically
AdS spacetime,
that DC-conductivities can be found by solving the Stokes-like equation on the black hole event horizon \cite{ban15,don15}. The non-vanishing
magnetic field in the theory in question was treated in \cite{don16}.

The obtained results revive the ones obtained by the concept of holographic Q-lattice used in studies of DC transport coefficients. Axion-like fields enable to break
the translation invariance, delivering the mechanism for momentum dissipation and leading to the finite value of DC coefficients. Interesting results using this technique
for various models and strength of dissipation were achieved  \cite{don14}-\cite{abb19a}. The studies in question were also elaborated in higher derivative gravity 
background \cite{don17}
and in Gauss-Bonnet-Maxwell scalar theory with momentum relaxation \cite{che15}.

%%%%%%%%%%%%%%%%%%%%%%%%%%%%%%%%%%%%%%%%%%%%%%%%

Our work will authorise the generalisation of the studies presented in \cite{don15, ban15} to describe CD thermoelectric conductivities in holographic Dirac semimetals, described on the gravity side by Einstein-Maxwell scalar gravity with auxiliary $U(1)$-gauge field coupled to the Maxwell one by the so-called {\it kinetic mixing} term.
Using foliation of the spacetime with hypersurfaces of constant $r$, we find the momentum constraints and derive the Stokes equations in terms of the linearised perturbation data
on the black brane event horizon. Solving these equations in the next step and finding the form of DC thermoelectric conductivity currents, we read off the adequate 
thermoelectric coefficients for the holographic Dirac semimetals. The principal objective of our studies is to reveal the influence of the auxiliary field, from the {\it hidden
sector} and $\alpha$-coupling constant on the thermoelectric conductivities in Dirac semimetals.

%%%%%%%%%%%%%%%%%%%%%%%%%%%%%%%%%%%%%%%%%%%%%%%%%
The paper is organised as follows. In section 2 we present the model of AdS Einstein-Maxwell scalar gravity with additional $U(1)$-gauge field
coupled to the ordinary one via the so-called {\it kinetic mixing term}. In section 3 the Arnowitt-Deser-Misner (ADM) formalism for the gravity model in question is elaborated.
We study the case of the spacetime foliation by hypersurfaces of constant $r$. Finding the momenta connected with the fields in the theory, by Legendre transformation,
we build the Hamiltonian and obtain the Hamilton equations on motion. Section 4 will be devoted to linear perturbations of the black brane spacetime and the
construction of gauge and heat currents, on the event horizon of the black object.
In section 5 we derive the Stokes equations on the black brane event horizon, for an incompressible doubly charged fluid.
Next one studies their main features. In section 6 we present the example of finding the CD thermoelectric conductivities obtained by solving
the linearised, time-independent, forced Stokes equations, in the case of one-dimensional Q-lattice.
We end up with summary and conclusions. 

%%%%%%%%%%%%%%%%%%%%%%%%%%%%%%%%%%%%%%%%%%%%%%%%%%%%%%%%%%%%%%%%%%%%%%%%%
%%%%%%%%%%%%%%%%%%%%%%%%%%%%%%%%%%%%%%%%%%%%%%%%%%%%%%%%%%%%%%%%%%%%%%%%
\section{Background holographic model}
In our paper we deal with the generalisation of the previously studied models \cite{don15,ban15}, by adding two interacting $U(1)$-gauge fields, in order to
find the influence of them on DC thermoelectric transport coefficients and to compare with the existing results.
In our model the gravitational action in $(3+1)$-dimensions is taken in the form  
\be
S = \int \sqrt{-g}~ d^4 x~  \bigg( R + \frac{6}{L^2} - \frac{1}{2} \na_\mu \phi \na^\mu \phi
- \frac{1}{4} F_{\mu \nu} F^{\mu \nu} - \frac{1}{4}B_{\mu \nu} B^{\mu \nu} - \frac{\alpha}{4} F_{\mu \nu} B^{\mu \nu} \bigg),
\label{sgrav} 
\ee
where $\phi$ is the scalar field, $F_{\mu \nu} = 2 \nabla_{[ \mu} A_{\nu ]}$ stands for the ordinary Maxwell field strength tensor, while
the second $U(1)$-gauge field $B_{\mu \nu}$ is given by $B_{\mu \nu} = 2 \nabla_{[ \mu} B_{\nu ]}$. $\alpha$ is a coupling constant between both gauge fields.
$L$ is the radius of AdS-spacetime. The coupling constant is denoted by $\alpha$.
%Predicted values of $\alpha$-coupling constant, being
%the kinetic mixing parameter between the two $U(1)$-gauge fields, for realistic string compactifications range between $10^{-2}$ and $10^{-16}$ \cite{abe04}-\cite{ban17}.

%%%%%%%%%%%%%%%%%%%%%%%%%%%%%%%
The justifications of such kind of gravity with electromagnetism coupled to the other gauge field exonerate from the top-down perspective \cite{ach16}. Namely, starting from
the string/M-theory the reduction to the lower dimensional gravity is performed. It relevant in the holographic correspondence attitude, because the theory 
in question is a fully consistent quantum theory (string/M-theory) and this fact 
 guarantees  that any predicted phenomenon by the top-down theory will be physical.
%%%%%%%%%%%%%%%%%%%%%%%%%%%%%%%%%%%%%%%

In the considered action (\ref{sgrav}) we have to do with the second gauge field coupled to the ordinary Maxwell one. This field is connected with the so-called {\it hidden sector} \cite{ach16}. The term describing interaction of visible (Maxwell field) sector and the {\it hidden} $U(1)$-gauge field is called the {\it kinetic mixing term}.
For the first time such types of terms were revealed in \cite{hol86}, in order to describe the existence and subsequent integrating out of heavy bi-fundamental fields charged under the $U(1)$-gauge groups. Predicted values of $\alpha$-coupling constant, being
the kinetic mixing parameter between the two $U(1)$-gauge fields, for realistic string compactifications range between $10^{-2}$ and $10^{-16}$ \cite{abe04}-\cite{ban17}.

The notion of the {\it hidden sector} stems from theories in which in addition to some visible gauge group we can encounter an
additional gauge group in the aforementioned sector. For instance, the compactified string or M-theory solutions generically result in
arising {\it hidden sectors}, which encompass at minimum the gauge fields and gauginos. It is caused by the various group factors contained in the gauge group symmetry of the
{\it hidden sector}. In the low-energy effective theory, the {\it hidden sector} comprehend states that are uncharged under the Standard Model symmetry group. 
On the contrary, they are charged under their own groups. On the other hand, {\it hidden sectors} interact with the visible ones via gravitational interaction. 
The interesting problem of the portals to our visible sector was discussed in \cite{portal1}-\cite{portal2}.

The presented model has its justification in string/M-theory, where the mixing portal (term which couples Maxwell and the additional $U(1)$-gauge field) arises
in open string theory. Gauge states are supported by D-branes separated in extra dimensions \cite{ach16}, in supersymmetric Type I, Type II A, Type II B models,
where the massive open strings stretch between two D-branes. The massive string/brane states existence connect the different gauge sectors.
The other cognisance of the above scenario can be achieved by 
M2-branes wrapped on surfaces which intersect two distinct codimension four orbifolds singularities. 
It happened that the natural generalisation can be carried out in M, F-theory and heterotic string models.

%%%%%%%%%%%%%%%%%%%%%%%%%%%%%
Many extensions of the Standard Model contain {\it hidden sectors} that have no renormalizable interactions with particle of the model in question. 
The $E8 \times E8$ string theory, being the realistic embeddings of the Standard Model and in type I, IIA, or IIB open string theory with branes, require the 
existence of the hidden sectors for the consistency and supersymmetry breaking \cite{abe08}. 
The most generic portal emerging from the string theory is the aforementioned {\it kinetic mixing} one.

The important fact is that the {\it kinetic mixing term} may contribute significantly and dominantly to the supersymmetry breaking mediation
\cite{abe04,die97},  proceeding in contributions to the scalar mass squared terms proportional to their hypercharges \cite{abe04}.

It happens that in string phenomenology \cite{abe08} the dimensionless {\it kinetic mixing term} parameter {$\alpha$} 
can be produced at an arbitrary high energy scale and it does not deteriorate from any kind of mass suppression from the messenger introducing it. 
Due to the fact that its measurement can provide some interesting features of 
high energy physics beyond the range of the contemporary colliders, this fact is o great significance for future experiments.

%%%%%%%%%%%%%%%%%%%%%%%%%%%%%%%%%%%%%%%%%%%%%%%%%%%%%%%%%%%%%%%%%%%%%%%%%%%%%%%%%%%%%%%%%%%%%%%%%%%

One also should remark that the model with two coupled vector fields, was used
in a generalisation of p-wave superconductivity, for the holographic model of ferromagnetic 
superconductivity \cite{amo14} and, without {\it kinetic mixing term}, for the description 
of the thermal conductivity in graphene \cite{seo17}. Various other settings of the model in question were described in \cite{nak14}-\cite{rog18b}, where among all it mimicked
{\it dark matter} sector influences on the properties of superconductors and superfluids, with or without taking additional assumptions about symmetries and chiral anomalies.

%%%%%%%%%%%%%%%%%%%%%%%%%%%%%%%%
Variation of the action $S$ with respect
to the metric, the scalar and gauge fields yields the following equations of motion:
\ben 
G_{\mu \nu} &-& g_{\mu \nu}~\frac{3}{L^2}  = T_{\mu \nu}(\phi) + T_{\mu \nu}(F) + T_{\mu \nu}(B) 
+ \alpha~T_{\mu \nu}(F,~B),\\ \label{ff1}
\na_{\mu}F^{\mu \nu} &+& \frac{\alpha}{2}~\na_\mu B^{\mu \nu} = 0,\\ \label{bb1}
\na_{\mu}B^{\mu \nu} &+& \frac{\alpha}{2}~\na_\mu F^{\mu \nu} = 0,\\
\na_\mu \na^\mu \phi &=& 0,
\een
where we have denoted by $G_{\mu \nu}$ the Einstein tensor, while the energy momentum tensors for the fields in the theory are given by
\ben
T_{\mu \nu} (\phi) &=& \frac{1}{2} \na_{\mu} \phi \na_\nu \phi - \frac{1}{4}~g_{\mu \nu}~\na_\delta \phi \na^\delta \phi ,\\
T_{\mu \nu}(F) &=& \frac{1}{2}~F_{\mu \delta}F_{\nu}{}^{\delta} - \frac{1}{8}~g_{\mu \nu}~F_{\alpha \beta}F^{\alpha \beta},\\
T_{\mu \nu}(B) &=& \frac{1}{2}~B_{\mu \delta}B_{\nu}{}^{\delta} - \frac{1}{8}~g_{\mu \nu}~B_{\alpha \beta}B^{\alpha \beta},\\
T_{\mu \nu}(F,~B) &=& \frac{1}{2}~F_{\mu \delta}B_{\nu}{}^{\delta} - \frac{1}{8}~g_{\mu \nu}~F_{\alpha \beta}B^{\alpha \beta}.
\een

For the gauge fields in the considered theory we assume the following  components:
\be
A_\mu ~dx^\mu = a_t ~dt, \qquad
B_\mu~dx^\mu  = b_t~dt.
\ee

%%%%%%%%%%%%%%%%%%%%%%%%%%%%%%%%%%%%%%%%%%%%%%%%%%%%%%%%%%%%%%%%%%%%%%%%%%%%%%
%%%%%%%%%%%%%%%%%%%%%%%%%%%%%%%%%%%%%%%%%%%%%%%%%%%%%%%%%%%%%%%%%%%%%%%%%%%%%%
\section{Arnowitt-Desser-Misner formalism }
In this section we present the basic idea for the $(3+1)$ formalism for the Einstein-Maxwell-scalar theory with additional $U(1)$-gauge field, coupled to the
ordinary Maxwell one. 
The formalism in question
considers the four-dimensional spacetime foliated
by three-geometries of constant $r$-coordinate. On three-dimensional hypersurfaces the induced metric may be written as
$h_{ab} = g_{ab} - n_a n_b$.
The line element is of the form as
\be
ds^{2} = N^2 dr^{2} + h_{ab} (dx^{a} + N^{a}dt)(dx^{b} + N^{b}dt),
\ee
where $N$ is the lapse function while $N^a$ stands for the shift vector for the constant $r$-coordinate hypersurface in the underlying manifold.
It turns out that the spacetime geometry can be described in terms of the intrinsic metric and the
extrinsic curvature of a three-dimensional hypersurface,
$N$ and $N^a$ relate the intrinsic
coordinate on one hypersurface to the intrinsic coordinates on a nearby
hypersurface. The general covariance allows the great arbitrariness in the choice of aforementioned functions
$N^\mu =  (N, ~N^{a})$.

In the canonical formulation of the Einstein-Maxwell scalar gravity with additional gauge field, 
the point in the phase space corresponds
to the specification of the fields
$(h_{ab}, ~\pi^{ab},~\tA_{i},~\tB_i,~\phi,~E_i,~B_i,~E)$ 
on a three-dimensional
$\Sigma_r$ manifold, where $h_{ab}$ denotes induced Riemannian metric on $\Sigma_r$,~
$\tA_i$ and $\tB_i$ are $U(1)$-gauge fields on 
the three-dimensional manifold, while $E_i$ and $B_i$ are respectively, Maxwell, auxiliary gauge {\it electric} fields in the evolved spacetime.
They constitute tensor densities quantities and imply the following relations:
\be
E_{k} = \sqrt{-h}~F_{\mu k} ~n^{\mu},
\label{}
\ee
for the Maxwell {\it electric} field,
while for the {\it dark matter} sector {\it electric} field $B_{i}$, one obtains
\be
B_{k} = \sqrt{-h}~B_{\mu k} ~n^{\mu}.
\label{}
\ee
ON the other hand, for the scalar field one has that 
\be
E = \sqrt{-h}~ \na_\mu \phi~n^{\mu},
\label{}
\ee
where in all the cases $n^{\mu}$ constitutes the unit normal vector to the hypersurface $\Sigma_r$ of constant $r$,
in the underlying spacetime. 

The Lagrangian density of Einstein double $U(1)$-gauge scalar gravity is subject to the relation
 \ben
 \cL = N \sqrt{-h} \Big[ {}^{(3)}R &-& K_{ab} K^{ab} + K^2  - \frac{1}{4} \Big( F_{ab} F^{ab}  + \frac{2}{(-h)} E_i E^i \Big) \\ \nonumber
 &-& \frac{1}{4} \Big( B_{ab} B^{ab}  + \frac{2}{(-h)} B_i B^i \Big)
 - \frac{\alpha}{4} \Big( F_{ab} B^{ab}  + \frac{2 \alpha}{(-h)} E_i B^i \Big) \Big],
 \een
where $K_{ab}$ denotes the extrinsic curvature, while ${}^{(3)}R $ is three-dimensional Ricci scalar.

In order to obtain 
the corresponding field momenta, one ought to perform variation of  the underlying Lagrangian
with respect to $\na_{r}h_{ab}$, ~$\na_{r}\pi_{ab}$,~ $\na_{r}\tA_{i}$,~$\na_{r}\tB_{i}$,
where $\na_{r}$ denotes the derivative with respect to $r$-coordinate. 

It leads respectively to the relation for gravitational momentum
\be
\pi^{ab} = - \sqrt{-h} \left ( K^{ab} - h^{ab}K \right ).
\label{}
\ee
Consequently, for the momentum responsible for $U(1)$-gauge fields one obtains
\be
\pi^{k}_{(F)} =\frac{\delta \cL}{\delta (\na_{r}\tA_{k})} = -E^k - \frac{\alpha }{2} B^k
, \qquad \pi^{k}_{(B)} = \frac{\delta \cL}{\delta (\na_{r}\tB_{i})} = - B^k - \frac{\alpha}{2} E^k,
\label{}
\ee
while for the scalar field $\phi$, it yields
\be
\pi_{(\phi)} =\frac{\delta \cL}{\delta(\na_r \phi)} = -E.
\ee

The Hamiltonian for the considered theory with auxiliary gauge field will be defined by the Legendre transform. It is given by
\ben \label{}
\cH &=& \pi^{ij} ~\na_{0} h_{ij} +  \pi^{i}_{(F)} ~\na_{0} \tA_{i} + \pi^{i}_{(B)} ~\na_{0} \tB_{i} 
- \cL \\ \nonumber
&=& N^{\mu} ~C_{\mu} + \tA_{r} ~{\tilde \tA} + \tB_{r} ~{\tilde \tB} + \cH_{div},
\een
where $\cH_{div}$ is the total derivative and has the form as follows:
\be
\cH_{div} = D_{k} \Big( \tA_r~ \pi^k_{(F)} + \tB_r~\pi^k_{(B)} \Big)
+ 2 D_{a} \Big( {N_{b} \pi^{ab} \over \sqrt{-h}} \Big),
\label{}
\ee
where $D_m$ denotes the covariant derivative with respect to the metric $h_{ab}$, on the hypersurface $\Sigma$.

%%%%%%%%%%%%%%%%%%%%%%%%%%%%%%%%%%%%%%%%%%%%%%%%%%
In our considerations we shall deal with the so-called {\it asymptotically flat initial data},
i.e., one has to do with
 an asymptotic region of hypersurface $\Sigma_r$, which is diffeomorphic
to ${\bf R}^3 -B$, where $B$ is compact. Moreover, the fall-off conditions for the fields in the phase space imply
\ben
h_{ab} &\approx& \delta_{ab} + \cO \Big( \frac{1}{r} \Big), \\
\pi_{ab} &\approx& \cO \Big( \frac{1}{r^2} \Big), \\
\tA_i &\approx& \cO \Big( \frac{1}{r} \Big), \qquad \tB_i \approx \cO \Big( \frac{1}{r} \Big),\\
E_i &\approx& \cO \Big( \frac{1}{r^2} \Big), \qquad B_i \approx \cO \Big( \frac{1}{r^2} \Big),\\
\phi &\approx& \cO \Big( \frac{1}{r^2} \Big), \qquad E \approx \cO \Big( \frac{1}{r^2} \Big).
\een
The standard asymptotic behaviour of the lapse function is provided by $N \approx 1 + \cO(1/r)$, while shift function behaves as
$N^a \approx \cO(1/r)$.
%%%%%%%%%%%%%%%%%%%%%%%%%%%%%%%%%%%%%%%%%%%%%%%%%%%%%%%%%%%%%%%%%%%%%%
Because of the fact that the r-component of $U(1)$ gauge fields $\tA_r,~\tB_r$ do not posses associated kinetic terms, we can consider them as
Lagrange multipliers, being subject to the generalised Gauss laws provided by
\be
{\tilde \tA} = - D_k \pi^k_{(F)} = 0, \qquad {\tilde \tB} = - D_k \pi^k_{(B)} = 0.
\ee
On the other hand, the components $C^\mu$ imply the relations as follows:
\ben \label{c0} \nonumber
C_0 &=& \frac{\sqrt{-h}}{4} \Big( F_{ab} F^{ab} + B_{ab}B^{ab} + \alpha ~F_{ab} B^{ab}  + 2 D_a \phi D^a \phi \Big) \\
&-& \frac{1}{2\sqrt{-h}} \Big( E_k E^k + B_k B^k +  \alpha~ E_k B^k  + E^2 \Big) \\ \nonumber
&+& \sqrt{-h} \Big( - {}^{(3)}R + \frac{6}{L^2}
-\Big( \frac{1}{(-h)}  \pi_{ij} \pi^{ij} - \frac{1}{2(-h)} \pi^2\Big) \Big), \\ \label{ca}
C_a &=& - \Big(  B_{ak} B^k + \frac{\alpha}{2} ~B_{ak} E^k + F_{ak} E^k + \frac{\alpha}{2} ~F_{ak} B^k  + D_a \phi E \Big)\\ \nonumber
&-& 2 \sqrt{-h}D_a \Big( \frac{N_b \pi^{ab}}{\sqrt{-h}} \Big).
\een
The evolution equations for the theory in question may be 
formally derived by considering the {\it volume integral contribution} to the Hamiltonian \cite{sud92},  denoted by $\cH_v$, which has pure constraint form, provided by
\be
\cH_v = \int_{\Sigma} d\Sigma~ N^\mu C_\mu.
\label{hv1}
\ee
Finding arbitrary infinitesimal variations $(\delta h_{ab}, ~\delta \pi_{ab},~\delta \tA_i,~\delta \tB_i,~\delta \phi,~\delta E_i,~\delta B_i,~\delta E)$, after integration by parts,
we obtain the change of the Hamiltonian $\cH_v$ caused by the variations in question, given by
\be
\delta \cH_v = \int_{\Sigma} d \Sigma  \Big(
P^{ab}~ \delta h_{ab} + Q^{ab} ~\delta \pi_{ab} + R^i ~\delta \tA_i + P^i ~\delta \tB_i + S^i ~\delta E_i + Q^i ~\delta B_i \Big).
\label{dhv}
\ee
The evolution equations for the considered system yield
\ben
{\Dot{ h}}_{ab} &=& \frac{\delta \cH_v}{\delta \pi^{ab} }= Q_{ab},\qquad
{\Dot{\pi}}_{ab} = - \frac{\delta \cH_v}{\delta h^{ab}}= - P_{ab}, \\
{\Dot{E}}_k &=& - \frac{\delta \cH_v}{\delta \tA^k} = -R_k, \qquad
{\Dot{\tA}}_k = \frac{\delta \cH_v}{\delta E^k } = S_k,\\
{\Dot{B}}_k &=& - \frac{\delta \cH_v}{\delta \tB^k } = -P_k, \qquad
{ \Dot{\tB}}_k = \frac{\delta \cH_v}{\delta B^k } = Q_k,\\
{\Dot{E}} &=& - \frac{\delta \cH_v}{\delta \phi } = -Z, \qquad
{ \Dot{\phi}} = \frac{\delta \cH_v}{\delta E } = W.
\een
%%%%%%%%%%%%%%%%%%%%%%%%%%%%%%%%%%%%%%%%%
where we have denoted
\ben \label{pab} \nonumber
P^{ab} &=& N \sqrt{-h}~ w^{ab} - N^m \Big( B_{m}{}^a B^b + \frac{\alpha}{2} B_m{}^a E^b + F_m{}^a E^b + \frac{\alpha}{2} F_m {}^a B^b \Big) + N^a ~D^b \phi \\
&+& \sqrt{-h} \Big( h^{ab} D^m D_m N - D^a D^b N \Big) 
- \cL_{N^i} \pi^{ab}, \\  \label{qab}
Q_{ab} &=& - \frac{N}{\sqrt{-h}} \Big( 2 \pi_{ab} - \pi_k{}^k~h_{ab} \Big) + \cL_{N^i} h_{ab},\\ \label{ri}
R^i &=& -\sqrt{-h} ~D_a \Big[ N \Big( F^{ai} + \frac{\alpha}{2} B^{ai}\Big) \Big] + \cL_{N^i} \Big( E^i + \frac{\alpha}{2} B^i \Big),\\ \label{pi}
P^i &=& -\sqrt{-h} ~D_a \Big[ N \Big( B^{ai} + \frac{\alpha}{2} F^{ai}\Big) \Big] + \cL_{N^i} \Big( B^i + \frac{\alpha}{2} E^i \Big),\\ \label{sk}
S_k &=& \frac{-N}{\sqrt{-h}} \Big( E_k + \frac{\alpha}{2} B_k \Big) - \Big( \cL_{N^i} \tA_k + \frac{\alpha}{2} \cL_{N^i} \tB_k \Big)
- D_k \Big(N \tA_r \Big) - \frac{ \alpha}{2}D_k \Big( N \tB_r \Big), \\ \label{qk}
Q_k &=& \frac{-N}{\sqrt{h}} \Big( B_k + \frac{\alpha}{2} E_k \Big) - \Big( \cL_{N^i} \tB_k + \frac{\alpha}{2} \cL_{N^i} \tA_k \Big)
- D_k \Big(N \tB_r \Big) + \frac{\alpha}{2} ~D_k \Big( N \tA_r \Big),\\
W &=& N^a ~D_a \phi,\\
Z &=& D_m \big( N^m~E \big).
\een
On the other hand, for the quantity $w^{ab}$,  entering (\ref{pab}), we have the following relation:
\ben \label{aab} \nonumber
w^{ab} &=&- \frac{1}{(-h)} \Big( 2 \pi^a{}_j \pi^{bj} - \pi_m{}^{m} \pi^{ab} \Big) + \frac{1}{2} h^{ab} \Big( \pi_{ij} \pi^{ij} - \frac{1}{2} \pi_m{}^m \pi_k{}^k \Big) 
+ {}^{(3)} R^{ab} - \frac{1}{2} h^{ab} {}^{(3)} R + \frac{3}{L^2}
\\ 
&+& 
\frac{h^{ab}}{8} \Big( F_{ij}F^{ij} + B_{ij}B^{ij} + \alpha F_{ij}B^{ij} + 2 D_m \phi D^m \phi \Big) \\ \nonumber
&+& \frac{1}{2} \Big(  F^{aj}B_j{}^b  + B^{aj}B_j{}^b + \alpha F^{aj}B_j{}^b + D^a \phi D^b \phi \Big) \\ \nonumber
&+& \frac{h^{ab}}{4 (-h)} \Big( E_k E^k + B_k B^k + \alpha E_k B^k + E^2 \Big) -
\frac{N}{2 (-h)} \Big( E^a E^b + B^a B^b + \alpha E^a B^b \Big).
\een
In the above formulae, $ \cL_{N^i}  E^k$, as well as $ \cL_{N^i}  B^k$ and $ \cL_{N^i}  \pi^{ab}$ denote the Lie derivatives of tensor densities and take the forms as
\ben
\cL_{N^i}  E^k &=& \sqrt{-h}~N^c D_c \Big(  \frac{E^k}{\sqrt{-h}} \Big) - E^c D_c N^k + E^k D_c N^c,\\
\cL_{N^i}  B^k &=& \sqrt{-h}~N^c D_c \Big(  \frac{B^k}{\sqrt{-h}} \Big) - B^c D_c N^k + B^k D_c N^c,\\
\cL_{N^i}  \pi^{ab} &=& \sqrt{-h}~N^c D_c \Big(  \frac{\pi^{ab}}{\sqrt{-h}} \Big) - 2 \pi^{c ( a} D_c N^{b)} + \pi^{ab} D_c N^c,
\een
while, $\cL_{N^i} \tA_k,~\cL_{N^i} \tB_k$ and $\cL_{N^i} h_{ab}$ correspond to the ordinary Lie derivatives.

The quantities $N^\mu,~\tA_0,~\tB_0$ are viewed as non-dynamical variables which are not represented in the phase space of
Einstein-Maxwel scalar auxiliary $U(1)$-gauge theory. It enables us to designate them arbitrarily.
The choice of $N^\mu$ is caused by the evolution of the considered system one looks for. On the other hand, $\tA_r$ and $\tB_r$ one restricts to the case
 when $r$-coordinates tends to infinity.

%%%%%%%%%%%%%%%%%%%%%%%%%%%%%%%%%%%%%%%%%%%%%%%%%%%%%%%%%%%%%%%%%%
%%%%%%%%%%%%%%%%%%%%%%%%%%%%%%%%%%%%%%%%%%%%%%%%%%%%%%%%%%%%%%%%%%
\section{Black brane spacetime}

In the our analysis  we  consider the  line element  of a charged under two $U(1)$-gauge groups, static black brane
\be
ds^2 = -U(r) G(r,x_i)dt^2 + \frac{F(r,x_i)dr^2}{U(r)} + ds^2(\Sigma_2),
\ee
where $\Sigma_2$ stands for the two-dimensional hypersurface at chosen $r$-coordinate. As in \cite{don15, ban15}, the line element at $r \rightarrow \infty$ approaches
the AdS boundary with the following conditions:
\ben
U \rightarrow r^2, \qquad F \rightarrow 1, \qquad G \rightarrow G(x), \qquad g_{ij} \rightarrow r^2~{\bar g}_{ij}\\
a_t(r, x_i) \rightarrow \mu(x), \qquad b_t(r, x_i) \rightarrow \mu_d(x), \qquad \phi(r, x_i) \rightarrow r^{\Delta -3} {\bar \phi}(x_i),
\een
where $\mu(x)$ and $\mu_d(x)$ are the spatially dependent chemical potentials bounded with the adequate $U(1)$-gauge field. 
${\bar \phi}(x_i)$ is a spatially dependent source for the accompanied with it dual operator. The operator has a dimensional scaling $\Delta$.
For the brevity of the subsequent notation we set the radius of AdS spacetime $L$ equal to one.

The black brane event horizon $\Sigma_2$, of defined topology, is situated at $r=0$. Having in mind the in-going coordinates
\be
v = t + \frac{\ln r}{4 \pi T} + \dots,
\ee
the near-horizon expansions of the metric tensor components and fields are given by \cite{ban15} the following relations:
\ben
U(r) &=& r \Big( 4 \pi T + U^{(1)} r + \dots \Big),\\
G(r,x_i) &=& G^{(0)}(x) + G^{(1)}(x) r + \dots, \\
F(r, x_i) &=& F^{(0)}(x) + F^{(1)} r + \dots, \\
g_{ij} &=& g^{(0)}_{ij} + g^{(1)}_{ij} r + \dots,\\
a_t(r, x_i) &=& r \Big( a^{(0)}_t ~G^{(0)}(x) + a^{(1)}_t(x) r + \dots \Big),\\
b_t(r, x_i) &=& r \Big( b^{(0)}_t ~G^{(0)}(x) + b^{(1)}_t(x) r + \dots \Big),\\
\phi(r, x_i) &=& \phi^{(0)}(x) + \phi^{(1)} (x) r + \dots,
\een
with the auxiliary condition that $G^{(0)}(x) = F^{(0)}(x)$.

%%%%%%%%%%%%%%%%%%%%%%%%%%%%%%%%%%%%%%%%%%%%%%%%%
%%%%%%%%%%%%%%%%%%%%%%%%%%%%%%%%%%%%%%%%%%%%%%%%%
\subsection{Perturbed black brane}
In the next step we turn on {\it electric} Maxwell $E_a$ and auxiliary $U(1)$-gauge $B_a$ fields, as well as, temperature gradient $\xi$,
in the spacetime under inspection, at fixed $r$-coordinate. The black brane will riposte to our action. We shall restrict our attention to
the linear appropriate perturbations $\delta g_{\mu \nu},~\delta a_\mu,~\delta b_\mu$, and $\delta \phi$.
The linear perturbations of the metric and fields are given by
\ben
\delta \Big(ds^2 \Big)  &=& \delta g_{\alpha \beta}~ dx^\alpha dx^\beta - 2 t~M~\xi_a dt~dx^a,\\
\delta A &=& \delta a_\beta ~dx^\beta - t~E_a dx^a + t~N~\xi_b~ dx^b,\\
\delta B &=& \delta b_\beta~ dx^\beta - t~B_a dx^a + t~N_d~\xi _b ~dx^b,
\een
as well as the perturbation of scalar field, $\delta \phi$. In what follows one supposes that
$\delta g_{\mu \nu},~\delta a_\mu,~\delta b_\mu$, and $\delta \phi$ are
functions of $(r,~x_m)$-coordinates. On the other hand, $E_a,~B_a, ~\xi_i$ are functions of $x_a$.
%%%%%%%%%%%%%%%%%%%%%%%%%%%%%%%%%
On the submanifold $\Sigma_2$ we demand that $E,~B,~\xi$ are closed one-forms, i.e.,
\be
d \Big( E_m dx^m \Big) = d \Big( B_m dx^m \Big) = d \Big( \xi_m dx^m \Big) = 0.
\ee
These assumptions are of great importance when the submanifold has torus topology or we consider $n$-dimensional black objects.
For instance studying five-dimensional spacetime, we have to consider torus topology of the event horizon for the so-called black ring, being stationary axisymmetric 
black object solution. The closed form assumption enables us to define potential and charges, i.e., calculating the Noether charges over the even horizon
we obtain constant value (potential) multiplied by the adequate charge \cite{cop05}-\cite{rog08}.
 Namely, using the Hodge theorem, the closed $p$-form, in $n$-dimensions, can be rewritten on the event horizon, as a sum of an exact and harmonic forms. An 
 exact form does not contribute to the equations because the equations of motion are satisfied. The harmonic part has the only contribution. The duality between 
 homology and cohomology concludes
 that there is a harmonic dual form to the $n-p-1$ cycle $S$, in the sense of the equality of the adequate surface integrals. Then it follows that the surface term will be of the form
 constant multiplied by the local charge.

%%%%%%%%%%%%%%
In order to establish the form of the linearised perturbations, one has to take into account that $t$-coordinate is no longer 
a good one at the black brane event horizon. The regularity of the perturbations near $r \rightarrow 0$, requires that some restrictions should be imposed
on them. Consequently, 
at the black brane near-horizon area, when $r \rightarrow 0$,  for the leading order, we obtain 
\ben
\delta g_{tt} &=& U(r) \Big( \delta g^{(0)}_{tt}(x_i) + \cO(r) \Big), \qquad  \delta g_{tr} = \delta g^{(0)}_{tr }(x_i) + \cO(r),\\
\delta g_{rr} &=& \frac{1}{U(r)} \Big( \delta g^{(0)}_{rr}(x_i) + \cO(r) \Big), \qquad \delta g_{ij} = \delta g^{(0)}_{ij}(x_i) + \cO(r),\\
\delta g_{ti} &=& \delta g^{(0)}_{ti}(x_i) - GU \xi_i \frac{\ln r}{4 \pi T} + \cO(r), \qquad \delta g_{ri} = \frac{1}{U(r)} \Big( \delta g^{(0)}_{ri} (x_i) + \cO(r) \Big),\\
\delta a_t &=& \delta a^{(0)}_t (x_i)+ \cO(r), \qquad \delta a_i = \frac{\ln r }{4 \pi T} \Big( - E_i + N \xi_i \Big) + \cO(r),\\
\delta a_r &=& \frac{1}{U(r)} \Big( \delta a^{(0)}_r(x_i) + \cO(r) \Big),\\
\delta b_t &=& \delta b^{(0)}_t (x_i)+ \cO(r), \qquad \delta b_i = \frac{\ln r }{4 \pi T} \Big( - B_i + N_d \xi_i \Big) + \cO(r),\\
\delta b_r &=& \frac{1}{U(r)} \Big( \delta b^{(0)}_r(x_i) + \cO(r) \Big),
\een
Moreover, it turns out that the constraint on the leading order have to be imposed. Namely one has that
\ben
\delta g^{(0)}_{tt} &+& \delta g^{(0)}_{rr} - 2 \delta g^{(0)}_{rt} = 0, \qquad \delta g^{(0)}_{ri} = \delta g^{(0)}_{ti},\\
\delta a^{(0)}_r &=& \delta a^{(0)}_t, \qquad \delta b^{(0)}_r = \delta b^{(0)}_t.
\een

%%%%%%%%%%%%%%%%%%%%%%%%%%%%%%%%%%%%%%%%%%%%%%%%%%%%%%%%%%%%%%%%%%%%%%%%%
\subsection{Electric currents}
In order to define the thermoelectric currents and heat conductivity one needs to find quantities in the bulk, which are identified with boundary currents.
We have to pay attention to the suitable Killing vector fields, as well as, the equations of motion to find the two-forms, being subject to the divergence of the
adequate coordinate equal to zero.

The {\it electric currents}  will be associated with the radially independent components 
of the equations (\ref{ff1}) and (\ref{bb1}), which in turn can be calculated everywhere
in the bulk. Because of the form the underlying equations they will constitute the mixture 
of the two $U(1)$-gauge fields. Consequently for the current connected with Maxwell gauge field we define
\be
J^i_{(F)}(r) = \sqrt{-g} ~\Big( F^{ir} + \frac{\alpha}{2} B^{ir} \Big),
\ee
while the current bounded with the auxiliary $U(1)$-gauge field yields
\be
J^i_{(B)}(r) = \sqrt{-g} ~\Big( B^{ir} + \frac{\alpha}{2} F^{ir} \Big).
\ee
On this account, having in mind relations for the background metric and components of the gauge fields, it is customary to write
\ben \nonumber \label{curf}
J^i_{(F)}(r) &=& \sqrt{G F g_d}~ g^{ij}_d ~ \Big[ \Big(
- \frac{\p_r a_t}{F G} \delta g_{tj} + \frac{\p_j a_t}{F G} \delta g_{tr} + \p_j (\delta a_r) \frac{U}{F} -
\Big( \p_r( \delta a_j) + t \p_r N~\xi_j \Big) \frac{U}{F} \Big) \\ 
&+& \frac{\alpha}{2}~\Big(
- \frac{\p_r b_t}{F G} \delta g_{tj} + \frac{\p_j b_t}{F G} \delta g_{tr} + \p_j (\delta b_r) \frac{U}{F} -
\Big( \p_r( \delta b_j) + t \p_r N_d~\xi_j \Big) \frac{U}{F} \Big) \Big].
\een
and for the other current 
\ben \nonumber \label{curb}
J^i_{(B)}(r) &=& \sqrt{G F g_d} ~g^{ij}_d ~ \Big[ \Big(
- \frac{\p_r b_t}{F G} \delta g_{tj} + \frac{\p_j b_t}{F G} \delta g_{tr} + \p_j (\delta b_r) \frac{U}{F} -
\Big( \p_r( \delta b_j) + t \p_r N_d~\xi_j \Big) \frac{U}{F} \Big) \\ 
&+& \frac{\alpha}{2}~\Big(
- \frac{\p_r a_t}{F G} \delta g_{tj} + \frac{\p_j a_t}{F G} \delta g_{tr} + \p_j (\delta a_r) \frac{U}{F} -
\Big( \p_r( \delta a_j) + t \p_r N~\xi_j \Big) \frac{U}{F} \Big) \Big].
\een
In the equations (\ref{curf}) and (\ref{curb}) we restrict our attention to the linearised order of black brane perturbations. Moreover the equations in question
envisage that no time-dependent terms are incorporated in them.

From the adequate equations of motion for $F_{\mu \nu}$ and $B_{\mu \nu}$ strength tensors, one can deduce that for Maxwell field current we have
the relation of the form 
\be
\na_i J^i_{(F)} = 0, \qquad \na_r J^i_{(F)} = \na_j \Big[ \sqrt{-g} \Big( F^{j i} + \frac{\alpha}{2} B^{ji} \Big) \Big],
\label{cfonh}
\ee
while for the auxiliary gauge, one achieves
\be
\na_i J^i_{(B)} = 0, \qquad \na_r J^i_{(F)} = \na_j \Big[ \sqrt{-g} \Big( B^{j i} + \frac{\alpha}{2} F^{ji} \Big) \Big].
\label{cbonh}
\ee

%%%%%%%%%%%%%%%%%%%%%%%%%%%%%%%%%%%%%%%%%%%%%%%%%%%%%%%%%%%%%%%%%%%%%%%%%
\subsection{Heat current}
In our set up we define the heat current
supposing that   $k_\mu = (\p/\p t)_\mu$ is a time-like Killing vector field \cite{rog18,rog18a}.  The general properties of Killing vector fields provide us the relation
\be
\na_\mu \na^\nu k^\mu = T^\nu {}{}_\mu k^\mu - \frac{k^\nu~T}{d-2} - 2 \frac{k^\nu \Lambda}{d-2},
\label{gen}
\ee
where $T =T_\mu {}{}^\mu$ is the trace of the energy momentum tensor in the considered theory while $\Lambda$ stands for cosmological constant.
The Killing vector symmetry conditions for the fields appearing in our model enable us to write
\be
\Lie_k F_{\alpha \beta} = \Lie_k B_{\alpha \beta}= \Lie_k \phi = 0.
\label{lie}
\ee
On the other hand, one has the following relations valid for arbitrary functions:
\be
k^\mu ~F_{\mu \nu} = \na_\nu \theta_{(F)}, \qquad k^\mu ~B_{\mu \nu} = \na_\nu \theta_{(B)},
\ee
where $\theta_{(F)}$ and  $\theta_{(B)}$ are arbitrary functions. By virtue of  (\ref{ff1}) and (\ref{bb1}) and the equation (\ref{lie}), one concludes that
\ben
k^\mu ~F_{\mu \alpha}F^{\rho \alpha} = \na_\alpha \Big( \theta_{(F)}~F^{\rho \alpha} \Big), \qquad
k^\mu ~B_{\mu \alpha}B^{\rho \alpha} = \na_\alpha \Big( \theta_{(B)}~B^{\rho \alpha} \Big),\\ 
k^\mu ~F_{\mu \ga}B^{\rho \ga} + k^\mu ~B_{\mu \alpha}F^{\rho \alpha} = \na_\delta \Big( \theta_{(F)} B^{\nu \delta} \Big) +  \na_\delta \Big( \theta_{(B)} F^{\nu \delta} \Big).
\een
Consequently with the above, we get the set of equations
\ben
k^\mu~F^{\rho \nu}F_{\rho \nu} &=& 4~\na_\rho \Big( k^{[\mu}F^{\rho]}A_\nu \Big) + 2~\Lie_k A_\nu~F^{\mu \nu}, \\
k^\mu~B^{\rho \nu}B_{\rho \nu} &=& 4~\na_\rho \Big( k^{[\mu}B^{\rho]}B_\nu \Big) + 2~\Lie_k B_\nu~B^{\mu \nu}, \\
k^\mu~B^{\rho \nu}F_{\rho \nu} &=& 4~\na_\rho \Big( k^{[\mu}F^{\rho]}B_\nu \Big) + 2~\Lie_k B_\nu~F^{\mu \nu}, \\
k^\mu~F^{\rho \nu}B_{\rho \nu} &=& 4~\na_\rho \Big( k^{[\mu}B^{\rho]}A_\nu \Big) + 2~\Lie_k A_\nu~B^{\mu \nu}, 
\een
After some algebra, having in mind the equation (\ref{gen}), it can be found that
\be
\na_\rho \tG_{\nu \rho} = - 2 \frac{\Lambda~k^\nu}{d-2},
\label{2form}
\ee
where the exact form of  $\tG_{\nu \rho} $ is given by
\ben \label{2forma}
\tG_{ \nu \rho} = \na^\nu k^\rho &+& \frac{1}{2} \Big( k^{[\nu}F^{\rho] \alpha}A_\alpha \Big) + \frac{1}{4} \Big[ \Big( \psi - 2 \theta_{(F)} \Big)~F^{\nu \rho} \Big] \\ \nonumber
&+& \frac{1}{2} \Big( k^{[\nu}B^{\rho] \alpha}B_\alpha \Big) + \frac{1}{4} \Big[ \Big( \chi - 2 \theta_{(B)} \Big)~B^{\nu \rho} \Big] \\ \nonumber
&+& \frac{\alpha}{4} \Big[ \Big( k^{[\nu}B^{\rho] \alpha}A_\alpha \Big) + \Big( k^{[\nu}F^{\rho] \alpha}B_\alpha \Big) \Big] \\ \nonumber
&+& \frac{\alpha}{8}\Big[ \Big( \psi - 2 \theta_{(F)} \Big)~B^{\nu \rho} \Big] + \frac{\alpha}{8} \Big[ \Big( \chi - 2 \theta_{(B)} \Big)~F^{\nu \rho} \Big].
\een
In the derivation of (\ref{2forma}) we have used the following relations:
\ben
\Lie_k A_\alpha~F^{\nu \alpha} = \na_\rho \Big( \psi~F^{\nu \rho} \Big), \qquad \Lie_k B_\alpha~B^{\nu \alpha} = \na_\rho \Big( \chi~B^{\nu \rho} \Big),\\
\Lie_k A_\alpha~B^{\nu \alpha} = \na_\rho \Big( \psi~B^{\nu \rho} \Big), \qquad \Lie_k B_\alpha~F^{\nu \alpha} = \na_\rho \Big( \chi~F^{\nu \rho} \Big),
\een
where we have denoted
\ben
\psi &=& E_\alpha x^\alpha, \qquad \theta_{(F)} = - E_\alpha x^\alpha - a_t,\\
\chi &=& B_\alpha x^\alpha, \qquad \theta_{(B)} = - B_\alpha x^\alpha - b_t.
\een
It can be observed that the $\tG_{\nu \rho}$ tensor is antisymmetric and implies
\be
\p_\rho \Big( 2~\sqrt{-g}~\tG^{\nu \rho} \Big) = - 2 \frac{\Lambda~\sqrt{-g}~k^\nu}{d-2}.
\ee
In what follows we shall use the two-form given by $2~\tG_{ \nu \rho} $, i.e.,
the heat current will be defined as $Q^i = 2~\sqrt{-g}~ \tG_{ \nu \rho} $.

%%%%%%%%%%%%%%%%%%%%%%%%%%%%%%%%%%%%%%%%%%%%%%%%%%%%%%%%%%%%%%%%%%%%%%%%%
Consequently, at the linear oder for the perturbed system in question, we obtain
\be
Q^i = \frac{G^{\frac{3}{2}}~U^2}{\sqrt{F}} \sqrt{g_d} ~g^{ij}_d ~ \Big[ \p_r \Big(\frac{\delta g_{tj}}{G U} \Big) - \p_j \Big( \frac{\delta g_{tr}}{G U} \Big) \Big]
- a_t J^i_{(F)} - b_t J^i_{(B)}.
\ee
As in the case of gauge currents, the time-dependent terms are dropped out of the expression for the heat current.
The additional relations for the heat current $Q^i$ are
\be
\na_i Q^i = 0, \qquad \na_r Q^i = \na_j \Big( 2~\sqrt{-g}~ \tG_{ \nu \rho} \Big).
\label{qonh}
\ee

%%%%%%%%%%%%%%%%%%%%%%%%%%%%%%%%%%%%%%%%%%%%%%%%%%%%%%%%%%%%%%%%%%%%%%%%%%%%%
\subsection{Currents on the black brane event horizon}
Accordingly to the equations (\ref{curf}) and (\ref{curb}), restricting our calculations to the linearised order of the perturbations we achieve the following currents on the 
black brane event horizon:
\ben
J^{i (0)}_{(F)} = J^I_{(F)} \mid_{\cH} &=& \sqrt{g^{(0)}} g^{ij}_{(0)}~
\Big[
\Big( \na_j (\delta a^{(0)}_t) + E_j - a^{(0)}_t ~\delta g^{(0)}_{tj} \Big) \\ \nonumber
&+& \frac{\alpha}{2} \Big( \na_j (\delta b^{(0)}_t) + B_j - b^{(0)}_t ~\delta g^{(0)}_{tj} \Big) \Big],\\
J^{i (0)}_{(B)} = J^I_{(B)} \mid_{\cH} &=& \sqrt{g^{(0)}} g^{ij}_{(0)}~
\Big[
\Big( \na_j (\delta b^{(0)}_t) + B_j - b^{(0)}_t ~\delta g^{(0)}_{tj} \Big) \\ \nonumber
&+& \frac{\alpha}{2} \Big( \na_j (\delta a^{(0)}_t) + E_j - a^{(0)}_t ~\delta g^{(0)}_{tj} \Big) \Big],\\
Q^{i (0)} = Q^{i} \mid_{\cH} &=& - 4 \pi~ T ~\sqrt{g^{(0)}} g^{ij}_{(0)}~\delta g^{(0)}_{tj}.
\een
By virtue of the equations (\ref{cfonh})-(\ref{cbonh}) and (\ref{qonh}), one has that for the perturbations at the black object horizon we get
\be
\na_i J^{i (0)}_{(F)} = 0, \qquad \na_i J^{i (0)}_{(B)} = 0, \qquad \na_i Q^{i (0)} = 0.
\ee

%%%%%%%%%%%%%%%%%%%%%%%%%%%%%%%%%%%%%%%%%%%%%%%%%%%%%%%%%%%%%%%%%%%%%%%%%%%%%%%
%%%%%%%%%%%%%%%%%%%%%%%%%%%%%%%%%%%%%%%%%%%%%%%%%%%%%%%%%%%%%%%%%%%%%%%%%%%%%%%
\section{Stokes equations for U(1)-gauge fluids}
In this section we obtain the closed system of differential equations which describe
the conditions imposed on a subset of a linearised perturbations, i.e., $\delta g^{(0)}_{it}, ~\delta g^{(0)}_{rt},~\delta a^{(0)}_t, ~\delta b^{(0)}_t,$
on the black brane event horizon.
On this account, it is customary to write
\ben \label{F}
\na_i \na^i w &+& \na_i E^i + \na_i \Big( a^{(0)}_t v^i \Big) + \frac{\alpha}{2} \Big[ \na_m \na^m w_d + \na_m B^m + \na_m \Big( b^{(0)}_t v^m \Big) \Big] = 0,\\ \label{B}
\na_i \na^i w_d &+& \na_i B^i + \na_i \Big( b^{(0)}_t v^i \Big) + \frac{\alpha}{2} \Big[ \na_m \na^m w + \na_m E^m + \na_m \Big( a^{(0)}_t v^m \Big) \Big] = 0,\\ \label{stokes}
b^{(0)}_t \Big[ \na_i w_d &+& B_i + \frac{\alpha}{2} \Big( \na_i w + E_i \Big) \Big]
+ a^{(0)}_t \Big[ \na_i w + E_i + \frac{\alpha}{2} \Big( \na_i w_d + B_i \Big) \Big] \\ \nonumber
&-& \na_i \phi^{(0)} \na_m \phi^{(0)} ~v^m + 2~\na^m \na_{(m} v_{i)} + 4 \pi T \xi_i - \na_i p = 0,\\ \label{vv}
\na_i v^i &=& 0,
\een
where we denoted
\ben \label{pp}
w &=& \delta a^{(0)}_t, \qquad w_d = \delta b^{(0)}_t, \qquad p = - 4 \pi T \frac{\delta g^{(0)}_{rt}}{G^{(0)}} - \delta g^{(0)}_{it} ~ \na^i \ln G^{(0)},\\
v_i &=& - \delta g^{(0)}_{it}.
\een
Calculating the conserved current equations, $ \na_i J^{i (0)}_{(F)} = 0$ and $\na_i J^{i (0)}_{(B)} = 0$, on the event horizon one can reach to the relations
(\ref{F})-(\ref{B}). On the other hand, when we take into account the constraints $C_i$ and $C_0$, we arrive respectively at equation (\ref{stokes}) and (\ref{vv}).
The covariant derivative $\na_i$  appearing in this section is bounded with the metric $g^{(0)}_{ij}$, on the black brane event horizon.
The relations (\ref{F})-(\ref{vv}) constitute a generalisation of the forced Stokes equations for charged $U(1)$-gauge fluids, described on the brane event horizon.

In the case of $a^{(0)}_t = b^{(0)}_t = w = w_d = E_i = B_i = 0$ and constant scalar field, one reaches the Stokes equations describing fluid with velocity $v_i$, pressure $p$,
and the additional forcing term of the form as $4 \pi T \xi_i$. As in the previously studied Einstein-Maxwell scalar case \cite{ban15}, we also have the viscosity factor given by
$ \na_i \phi^{(0)} \na_m \phi^{(0)}~ v^m$.

It can be remarked that taking the divergence of the equation (\ref{stokes}) and having in mind the remaining relations, we arrive at the pressure Poisson equation,
being the generalisation of the one derived in \cite{ban15}. Namely, it is provided by the following expression:
\ben \nonumber
\na^m \na_m p &=&
\na_j \Big[ b^{(0)}_t \Big( \na^j w_d + B^j + \frac{\alpha}{2} \Big(\na^j w + E^j  \Big) \Big)
+ a^{(0)}_t \Big( \na^j w + E^j + \frac{\alpha}{2} \Big(\na^j w_d + B^j \Big) \Big) \\
&-&\na^j  \phi^{(0)} \na_m \phi^{(0)} v^m + 4 \pi T \xi^j + 2 R^j{}{}_m v^m.
\een
On the other hand, multiplication of (\ref{stokes}) and integration of the resulting expression over the black brane event horizon reveal
\ben \nonumber \label{pos}
\int \sqrt{g^{(0)}} d^2 x \Big[ 
 2 \na^{(i}v^{j)} \na_{(i}v_{j)} &+& \Big( \na_i w + E_i \Big)\Big( \na^i w + E^i \Big) + \Big( \na_i w_d + B_i \Big)\Big( \na^i w_d + B^i \Big)\\
 \alpha \Big( \na_i w &+& E_i \Big)\Big( \na^i w_d + B^i \Big)
+ v^m \na_m \phi^{(0)} \na_j \phi^{(0)} v^j \Big] \\ \nonumber
&=& \int d^2x \Big[ Q^{i (0)}  \xi_i + J^ {i (0)}_{(F)}  E_i + J^ {i (0)}_{(B)}  B_i \Big].
\een
Having in mind the non-compactness of the event horizon and assuming the disappearance of the emerging surface terms, the above equation reveals
the fact that its left-hand side is positive, which in turn implies the positivity of the thermoelectrical conductivities bounded with the two $U(1)$-gauge and scalar fields,
emerging on the right-hand side of (\ref{pos}).

As far as the uniqueness of the above set of equations (\ref{F})-(\ref{vv}) is concerned, let us suppose that we have two solutions of this set subject to the same boundary and regularity conditions. The differences of them will be denoted as ${\tilde v}_i = v^{(1)}_i - v^{(2)}_i,~{\tilde w} = w^{(1)}- w^{(2)},~ {\tilde w}_d = w^{(1)}_d- w^{(2)}_d,~ 
{\tilde p} = p^{(1)}-p^{(2)}$. They will satisfy the equation with $\xi_i = E_i = B_i =0$. Just using the relation (\ref{pos}) one obtains
\be
 \na_{(i}{\tilde v}_{j)} =0, \qquad \na_i {\tilde w} = 0, \qquad  \na_i {\tilde w}_d = 0, \qquad {\tilde v}^i \na_i \phi^{(0)} = 0.
 \ee
Equations (\ref{F}) and (\ref{B}) reveal that the Lie derivatives with respect to $v_m$ taken from $a^{(0)}_t$ and $b^{(0)}_t$ are equal to zero, i.e.,
\be 
\cL_{v^m} a^{(0)}_t = 0, \qquad \cL_{v^m} b^{(0)}_t = 0,
\ee
while the relation (\ref{}) leads to the condition $\na_i {\tilde p} = 0$. Having all these in mind 
one may conclude that the solution of Stokes equation is unique up to the Killing vectors of the
black brane event horizon line element, with $\tilde p,~\tilde w, ~{\tilde w}_d$ constant and $\delta g^{(0)}_{rt}$ described by the equation (\ref{pp}).

The aforementioned set of equation can be derived by varying the following functional:
\ben \label{var} \nonumber
S &=& \int \sqrt{g^{(0)}} d^4x \Bigg[
- \na ^{(i}v^{j)} \na_{(i} v_{j)} + 4 \pi T \xi_m v^m  - \frac{1}{2} v_i \na^i \phi^{(0)}~ v_k \na^k \phi^{(0)} + p~\na_m v^m \\ 
&+& E_i \Big( a^{(0)}_t v^i + \na^i w \Big) + \frac{1}{2} E_m E^m + \frac{\alpha}{2} E_i \Big( b^{(0)}_t v^i + \na^i w_d \Big) + \frac{\alpha}{2} E_k B^k \\ \nonumber
&+& B_i \Big( b^{(0)}_t v^i + \na^i w_d \Big) + \frac{1}{2} B_m B^m + \frac{\alpha}{2} B_i \Big( a^{(0)}_t v^i + \na^i w \Big) \\ \nonumber
&+& \frac{1}{2} \Big( b^{(0)}_t v^i + \na^i w_d \Big)\Big( b^{(0)}_t v_i + \na_i w_d \Big) - \frac{1}{2} {b^{(0)}_t}^2 v_m v^m \\ \nonumber
&+& \frac{1}{2} \Big( a^{(0)}_t v^i + \na^i w \Big)\Big( a^{(0)}_t v_i + \na_i w \Big) - \frac{1}{2} {a^{(0)}_t}^2 v_m v^m \\ \nonumber
&+& \frac{\alpha}{2} \Big[ \frac{1}{2} \Big( b^{(0)}_t v^i + \na^i w \Big)\Big( b^{(0)}_t v_i + \na_i w_d \Big) - \frac{1}{2} {b^{(0)}_t}^2 v_m v^m \Big] \\ \nonumber
&+& \frac{\alpha}{2} \Big[ \frac{1}{2} \Big( a^{(0)}_t v^i + \na^i w_d \Big)\Big( a^{(0)}_t v_i + \na_i w_d \Big) - \frac{1}{2} {a^{(0)}_t}^2 v_m v^m \Big] \Bigg].
\een

It happens that the pressure appears here as the Lagrange multiplier, giving the incompressibility condition. The rest of the relations in question
can be found by varying with respect to $v_i, ~w$ and $w_d$. Moreover, variations of (\ref{var}) with respect to $E_i, ~B_i$ and $\xi_i$ give us appropriate
currents of gauge fields and heat, counted on the black brane event horizon.

Further, having in mind the fact that $E_i,~B_i,~\xi_i$ can be described as closed differential forms, and consequently they are locally defined as
\be
E = \na_m e ~dx^m, \qquad B = \na_m b~dx^m, \qquad \xi = \na_m z~dx^m,
\ee
the studied differential system of equation can be rewritten in order to eliminate source terms. Namely, defining the quantities
\be
{\tilde w} = w + e, \qquad {\tilde w}_d = w_d + b, \qquad {\tilde p} = p - 4 \pi T z,
\ee
the equations (\ref{F})-(\ref{vv}) have the forms as follows:
\ben \label{tf}
\na_i \na^i {\tilde w} &+& \na_i \Big( a^{(0)}_t v^i \Big) + \frac{\alpha}{2} \Big[ \na_m \na^m {\tilde w}_d + \na_m \Big( b^{(0)}_t v^m \Big) \Big] = 0,\\ \label{tb}
\na_i \na^i {\tilde w}_d &+& \na_i \Big( b^{(0)}_t v^i \Big) + \frac{\alpha}{2} \Big[ \na_m \na^m {\tilde w} + \na_m \Big( a^{(0)}_t v^m \Big) \Big] = 0,\\ \label{tstokes}
b^{(0)}_t \Big( \na_i {\tilde w}_d &+&  \frac{\alpha}{2} \na_i {\tilde w}  \Big)
+ a^{(0)}_t \Big( \na_i {\tilde w} + \frac{\alpha}{2}  \na_i {\tilde w}_d  \Big) \\ \nonumber
&-& \na_i \phi^{(0)} \na_m \phi^{(0)} ~v^m + 2~\na^m \na_{(m} v_{i)}  - \na_i {\tilde p} = 0,\\ \label{tvv}
\na_i v^i &=& 0.
\een

%%%%%%%%%%%%%%%%%%%%%%%%%%%%%%%%%%%%%%%%%%%%%%%%%%%%%%%%%%%%%%%%
%%%%%%%%%%%%%%%%%%%%%%%%%%%%%%%%%%%%%%%%%%%%%%%%%%%%%%%%%%%%%%%%%
\section{One-dimensional Q-lattice case}
The purpose of this section is to elaborate the example of holographic lattice for which one solves the previously derived Stokes equations and finds the DC thermoelectric
conductivities. We shall be interested in the influence of the field from the {\it hidden sector}, as well as, the dependence of the conductivities
on $\alpha$-coupling constant appearing in the {\it mixing term}, on the physics of this phenomena.

In what follows we shall study the line element of black brane which breaks the spatial translation symmetry in one-dimension, on the black object
event horizon. Namely, let us suppose that the event horizon metric tensor depends only on $x$-coordinate and the 
two-dimensional line element implies
\be
ds_{(2)}^2 = \ga(x) dx^2 + \la(x) dy^2.
\ee
In the next step one tries to solve equations (\ref{stokes})-(\ref{vv}) in the aforementioned background. The incompressibility condition described by (\ref{vv})
yields the following relation:
\be
v^i = \frac{1}{\sqrt{ \ga(x) \la(x)}} v_0,
\ee
where $v_0$ is constant.

On the other hand, the Stokes equation can be rewritten in the form as
\ben
\frac{1}{\sqrt{g^{(0)}}} \Big( b^{(0)}_t ~J^{i(0)}_{(B)} &+& a^{(0)}_t~J^{i(0)}_{(F)} \Big)
- v^i ~\Big( a^{(0) 2}_t + \alpha ~a^{(0)}_t~ b^{(0)}_t + b^{(0) 2}_t \Big) \\ \nonumber
&-& \na^i \phi^{(0)} \na_m \phi^{(0)} v^m + 2 \na_m \na^{(m} v^{i)} - \na^i p = - 4 \pi T \xi^i
 \een
The Stokes equation and the relation describing currents for the two $U(1)$-gauge fields, are implemented to describe the functions $w,~w_d$ and $p$.
As we consider the periodic functions, one has to have that the expressions for $\p_x w,~\p_x w_d,~\p_x p$ have no zero modes on the considered submanifold.
In turn this requirement imposes the constraints on $J^{x (0)}_{(F)},~J^{x (0)}_{(B)}$ and $v_0$. As in \cite{ban15}, we find these restrictions by averaging the equations
over the periodic lattice.

Before proceeding to this task let us find the exact forms of the gauge currents and the Stokes equation in the two-dimensional submanifold.
The $U(1)$-gauge currents are provided by the following relations:
\ben
J^{x (0)}_{(F)} &=& \sqrt{\frac{\la(x)}{\ga(x)}} \Big[
\Big( \p_x w + E_x + \sqrt{\frac{\ga(x)}{\la(x)}} a^{(0)}_t ~v_0 \Big)
+ \frac{\alpha}{2} \Big( \p_x w_d + B_x + \sqrt{\frac{\ga(x)}{\la(x)}} b^{(0)}_t ~v_0 \Big), \\
J^{x (0)}_{(B)} &=& \sqrt{\frac{\la(x)}{\ga(x)}} \Big[
\Big( \p_x w_d + B_x + \sqrt{\frac{\ga(x)}{\la(x)}} b^{(0)}_t ~v_0 \Big)
+ \frac{\alpha}{2} \Big( \p_x w + E_x + \sqrt{\frac{\ga(x)}{\la(x)}} a^{(0)}_t ~v_0 \Big),
\een
while the heat current implies
\be
Q^{i (0)} = 4 \pi T~v_0.
\label{qo}
\ee
Consequently the Stokes equation can be rewritten as
\be
v_0~\p_x \Big( \ga^{-\frac{1}{2}} \p_x ( \la^{-\frac{1}{2}})  \Big) - Y~v_0 + \sqrt{\frac{\ga(x)}{\la(x)}} \Big(  b^{(0)}_t ~J^{i(0)}_{(B)} + a^{(0)}_t~J^{i(0)}_{(F)} \Big)
-\p_x p = - 4 \pi T~\xi_x,
\label{onestok}
\ee
where one denotes by $Y$ the quantity which yields
\be
Y = \frac{(\p_x \phi)^2}{\sqrt{\la(x) \ga(x)}} +  a^{(0)2}_t + \alpha~a^{(0)}_t b^{(0)}_t + b^{(0)2}_t + \frac{1}{4 { \la (x)}^{\frac{5}{2}} {\ga(x)}^{\frac{1}{2}}}
{( \p_x \la (x))}^2
 v_0.
\ee
Averaging the Stokes equation (\ref{onestok}) over one-dimensional lattice with a period $L_1$, leads to the following:
\ben \nonumber \label{vo}
v_0 &=& \frac{1}{A} \Big[ 
\Big( B_x + \frac{\alpha}{2} E_x \Big) \int dx \sqrt{\frac{\ga(x)}{\la(x)}} ~b^{(0)}_t + \Big( E_x + \frac{\alpha}{2} B_x \Big) \int dx \sqrt{\frac{\ga(x)}{\la(x)}} ~a^{(0)}_t \\
&+& 4 \pi T~\xi_x ~\int dx \sqrt{\frac{\ga(x)}{\la(x)}} \Big],
\een
where one sets
\ben
A = \int dx \sqrt{\frac{\ga(x)}{\la(x)}} ~\int dx ~Y &-& \Big( \int dx \sqrt{\frac{\ga(x)}{\la(x)}}~a^{(0)}_t \Big)^2  - \Big( \int dx \sqrt{\frac{\ga(x)}{\la(x)}}~b^{(0)}_t \Big)^2 \\
\nonumber
 &-& \alpha~\Big( \int dx \sqrt{\frac{\ga(x)}{\la(x)}}~a^{(0)}_t \Big)~\Big( \int dx \sqrt{\frac{\ga(x)}{\la(x)}}~b^{(0)}_t \Big),
  \een
where for the brevity of the future notion we set
\be
\int dx \leftrightarrow \frac{1}{L_1} \int_0^{L_1}dx.
\ee

By virtue of the same procedure that we followed above, we arrive at the relations for Maxwell current $J^{x (0)}_F$ 
\ben \label{jf}
J^{x (0)}_{(F)}
&=& \frac{1}{A} \Bigg[
\Big(E_x + \frac{\alpha}{2} B_x \Big) \Big[ \int dx ~Y - \int dx \sqrt{\frac{\ga(x)}{\la(x)}}~b^{(0)}_t \Big( b^{(0)}_t + \frac{\alpha}{2} a^{(0)}_t \Big) \Big] \\ \nonumber
&+& 4 \pi T~\xi_x \int dx \sqrt{\frac{\ga(x)}{\la(x)}}~\Big( a^{(0)}_t + \frac{\alpha}{2} b^{(0)}_t \Big) 
+ \Big(B_x + \frac{\alpha}{2} E_x \Big) \int dx \sqrt{\frac{\ga(x)}{\la(x)}}~b^{(0)}_t \Big( a^{(0)}_t + \frac{\alpha}{2} b^{(0)}_t \Big)  \Bigg],
\een
and respectively for $J^{x (0)}_B$
\ben \label{jb}
J^{x (0)}_{(B)} &=& \frac{1}{A} \Bigg[
\Big(B_x + \frac{\alpha}{2} E_x \Big) \Big[ \int dx ~Y - \int dx \sqrt{\frac{\ga(x)}{\la(x)}}~a^{(0)}_t \Big( a^{(0)}_t + \frac{\alpha}{2} b^{(0)}_t \Big) \Big] \\ \nonumber
&+& 4 \pi T~\xi_x \int dx \sqrt{\frac{\ga(x)}{\la(x)}}~\Big( b^{(0)}_t + \frac{\alpha}{2} a^{(0)}_t \Big) 
+ \Big(E_x + \frac{\alpha}{2} B_x \Big) \int dx \sqrt{\frac{\ga(x)}{\la(x)}}~a^{(0)}_t \Big( b^{(0)}_t + \frac{\alpha}{2} a^{(0)}_t \Big)  \Bigg].
\een
For the heat current one finds the relation as in equation (\ref{qo}), where $v_0$ is given by the equation (\ref{vo}).

It can be pointed out that in the limit when $\alpha =0,~b^{(0)}_t =0,~ B_x =0$, one receives the relations presented in \cite{ban15}, for the case of Einstein-Maxwell scalar
theory.

%%%%%%%%%%%%%%%%%%%%%%%%%%%%%%%%%%%%%%%%%%

\subsection{Kinetic and transport coefficients in one-dimensional case}
In our model, the general form of the kinetic coefficient matrix may be written as
\begin{eqnarray}
& \left(
\begin{array}{c}
{J_{(F)}^{i (0) }}\\
{J_{(B)}^{i (0)}}\\
{Q^{i (0)}}\\
\end{array}
\right)
=
& \left(\begin{array}{ccc}
 {\sigma{^i_j}_{(FF)}} &  {\sigma{^i_j}_{(FB)}} & {\alpha{^i_j}_{(F)}}T  \\
 {\sigma{^i_j}_{(BF)}} & {\sigma{^i_j}_{(BB)}} &{\alpha{^i_j}_{(B)}}T  \\
 {\alpha{^i_j}_{(F)}}T & {\alpha{^i_j}_{(B)}}T &\ {\kappa{^i_j}}T  \\
 \end{array}
 \right)
\left(
\begin{array}{c}
{E^j}\\
 {B^j}\\
 {\xi^j}\\
\end{array}
\right).
\label{ji-vs-fieldsi}
 \end{eqnarray}
%with $i,j=x,y$ and obvious definitions $\sigma_{(ab)}^{ij}=\frac{\partial J^i_{(a)}}{\partial E_{j (b)}}$ 
%of various conductances $\sigma_{(ab)}^{ij}$ with $a,~b=F,~B$, thermoelectric components
%$\alpha_{(a)}^{ij}=\frac{\partial J^i_{(a)}}{\partial \xi_j}$ and $\kappa_{0}^{ij}=\frac{\partial \tQ^i}{\partial \xi_j}$.
The relation (\ref{ji-vs-fieldsi}) will hepl us to find their exact form for one-dimensional Q-lattice case. Namely,
combining the equations (\ref{qo}),~(\ref{vo}),~ (\ref{jf})-(\ref{jb}) and using the relation (\ref{ji-vs-fieldsi}), we obtain the required explicit form of the kinetic
coefficients in the case of one-dimensional lattice. They are provided by
\ben \label{sff}
\sigma_{FF} &=& \frac{1}{A} \Big[
\int dx~ Y - \talpha \int dx \sqrt{\frac{\ga(x)}{\la(x)}}~b^{(0) 2}_t \Big], \\
\sigma_{FB} &=&  \frac{1}{A} \Big[ \frac{\alpha}{2}
\int dx~ Y + \talpha \int dx \sqrt{\frac{\ga(x)}{\la(x)}}~a^{(0)}_t~b^{(0) }_t \Big], \\
\sigma_{BF} &=&  \frac{1}{A} \Big[ \frac{\alpha}{2}
\int dx~ Y + \talpha \int dx \sqrt{\frac{\ga(x)}{\la(x)}}~a^{(0)}_t ~b^{(0) }_t \Big], \\
\sigma_{BB} &=&  \frac{1}{A} \Big[
\int dx ~Y - \talpha \int dx \sqrt{\frac{\ga(x)}{\la(x)}}~a^{(0) 2}_t \Big], \\
\alpha_{(F)} &=& \frac{4 \pi}{A} \int dx \sqrt{\frac{\ga(x)}{\la(x)}}
~\Big( a^{(0)}_t + \frac{\alpha}{2} b^{(0)}_t \Big),\\
\alpha_{(B)} &=& \frac{4 \pi}{A} \int dx \sqrt{\frac{\ga(x)}{\la(x)}}
~\Big( b^{(0)}_t + \frac{\alpha}{2} a^{(0)}_t \Big), \\ \label{ka}
\kappa &=& \frac{(4 \pi)^2 T}{A} \int dx \sqrt{\frac{\ga(x)}{\la(x)}},
\een
where $\talpha = 1 - \frac{\alpha^2}{4}$.

The form of the kinetic and transport coefficients (\ref{sff})-(\ref{ka}) envisage the strong influence of the auxiliary $U(1)$-gauge field and the coupling constant $\alpha$
on their structure.

%%%%%%%%%%%%%%%%%%%%%%%%%%%%%%%%%%%%%%%%%%%%%%%%%%%%%
\subsection{Generalisation of the Sachdev model of the Dirac fluid}
The analysis presented in this subsection is addressed the the question about
the connections of the described model with the results presented in \cite{seo17}, where
the holographic two current model of Dirac fluid in graphene, has been presented. On the other hand, the
generalisation for the case of two interacting $U(1)$-gauge
currents were performed, for holographic four \cite{rog18} and five-dimensional models \cite{rog18a}.

In this subsection we shall find how the aforementioned results can be found in the Stokes equations.
We suppose, for simplicity, that the submanifold will be ${\bf R}^2$ one with flat metric described by
line element $ds^2_{(2)}$. We set that $\gamma(x) = \la(x) =1$.

The total electric current will be the sum of the {\it visible } and {\it hidden} sector currents 
\be
J^{x (0)} = J^{x (0)}_{(F)} + J^{x (0)}_{(B)},
\ee
while the electric conductivity is provided by
\be
\sigma = \sigma_{FF} + \sigma_{FB} + \sigma_{BF} + \sigma_{BB}.
\ee

On the other hand, the gauge and heat currents are provided by
\ben \nonumber
J^{x (0)}_{(F)} &=& E_x~\frac{1}{A} \int dx \Big[ (\p_x \phi)^2 + Q_F^2 \Big] + B_x \frac{1}{A} \int dx \Big[ \frac{\alpha}{2}(\p_x \phi)^2 + Q_F~Q_B \Big] \\ 
&+& 4 \pi~T \xi_x ~\frac{1}{A} \int dx~Q_F,\\
J^{x (0)}_{(B)} &=& B_x~\frac{1}{A} \int dx \Big[ (\p_x \phi)^2 + Q_B^2 \Big] + E_x \frac{1}{A} \int dx \Big[ \frac{\alpha}{2}(\p_x \phi)^2 + Q_F~Q_B \Big] \\ 
&+& 4 \pi~T \xi_x ~\frac{1}{A} \int dx~Q_B,\\
Q^{x (0)} &=& 4 \pi~T E_x~\frac{1}{A} \int dx~Q_F + 4 \pi~T B_x~\frac{1}{A} \int dx~Q_B + \Big( 4 \pi T \Big)^2~ \frac{1}{A} \int dx,
\een
where in order to have the more transparent connection with the results presented in \cite{rog18,rog18a}, we set
\ben
Q_F &=&  a^{(0)}_t + \frac{\alpha}{2} b^{(0)}_t ,\\
Q_B &=& b^{(0)}_t + \frac{\alpha}{2} a^{(0)}_t.
\een
As in \cite{rog18}, let us suppose that the two charges will be connected by the relation
\be
b^{(0)}_t = g~a^{(0)}_t,
\ee
where $g$ is a number. The definitions of $Q_F$ and $Q_B$ enable us to find that
\be
Q_F = \Big( 1 + \frac{\alpha}{2} g \Big)~a^{(0)}_t, \qquad Q_B = \Big( g + \frac{\alpha}{2} g \Big)~a^{(0)}_t.
\ee

Consequently the electrical conductivity is given by
\be
\sigma = \frac{1}{A} \int dx~\Big[ \Big( 1 + \alpha \Big) (\p_x \phi)^2 + \Big( 1 + \frac{\alpha}{2} \Big)^2 \Big( 1 + g \Big)^2~a^{(0) 2}_t \Big].
\ee
If we define $Q = Q_F + Q_B$, the above relation for $\sigma$ reduces to the form as
\be
\sigma = \frac{1}{A} \int dx~\Big[ \Big( 1 + \alpha \Big) (\p_x \phi)^2 + Q^2 \Big].
\ee
One can see that we have a very good agreement with the results achieved in \cite{rog18} and what it is more in the Stokes equations attitude, one has the explicit
presence of the scalar field (influenced on the the electric conductivity) responsible for the dissipation processes. It is given by the square 
of its derivative with respect to the spatial coordinate $x$.

%%%%%%%%%%%%%%%%%%%%%%%%%%%%%%%%%%%%%%%%%%%%%%%%%%%%%%%%%%%%%%%%%%%%%%%%%%%%%%%%%%%%%
\subsubsection{Absence of {\it hidden sector} fields}
To proceed further,
in this subsection we shall elaborate the specific choice of matter fields for the one-dimensional periodic lattice.
Our main aim will be to find kinetic coefficients generated by the aforementioned matter fields.
To commence with we shall choose the case of the nonexistence of {\it hidden sector} fields,
i.e., ${\alpha =0,~b^{(0) }_t=0, B_x =0}$. One has that for this definite option, we obtain
the currents are equal to the following expressions:
\ben \label{fvis}
J^{x (0)}_{(F)} &=& \frac{1}{A_{vis}} \Big[
E_x ~\int dx ~Y_{vis} + 4 \pi T~\xi_x \int dx \sqrt{\frac{\ga(x)}{\la(x)}}~ a^{(0)}_t \Big], \\ \label{bvis}
J^{x (0)}_{(B)} &=& 0.
\een
where one sets
\be
v_0 (vis) = \frac{1}{A_{vis}} \Big[ E_x  \int dx \sqrt{\frac{\ga(x)}{\la(x)}} ~a^{(0)}_t \\
+ 4 \pi T~\xi_x ~\int dx \sqrt{\frac{\ga(x)}{\la(x)}} \Big],
\ee
\be
A_{vis} = \int dx \sqrt{\frac{\ga(x)}{\la(x)}} ~\int dx ~Y_{vis} - \Big( \int dx \sqrt{\frac{\ga(x)}{\la(x)}}~a^{(0)}_t \Big)^2,
 \ee
\be
Y_{vis} = \frac{(\p_x \phi)^2}{\sqrt{\la(x) \ga(x)}} +  a^{(0)2}_t + \frac{1}{4 { \la (x)}^{\frac{5}{2}} {\ga(x)}^{\frac{1}{2}}}
{( \p_x \la (x))}^2~
 v_0(vis).
\ee
A comparison of the equation (\ref{ji-vs-fieldsi}), ~(\ref{fvis})-(\ref{bvis}) reveals the DC thermoelectric coefficients as follows:

We obtain the following kinetic coefficients:
\ben \label{vis}
\sigma_{FF} &=& \frac{1}{A_{vis}} \int dx~ Y_{vis}, \\
\sigma_{FB} &=&  \sigma_{BF} = 0,\\
\sigma_{BB} &=& 0, \\
\alpha_{(F)} &=& \frac{4 \pi}{A_{vis}} \int dx \sqrt{\frac{\ga(x)}{\la(x)}}
~a^{(0)}_t,\\
\alpha_{(B)} &=& 0,\\
\kappa &=& \frac{(4 \pi)^2 T}{A_{vis}} \int dx \sqrt{\frac{\ga(x)}{\la(x)}},
\een
As it should be expected, we have obtained the results presented in \cite{ban15} for Einstein-Maxwell scalar gravity with non-zero potential.
We have reached the anticipated limit of our model.
%%%%%%%%%%%%%%%%%%%%%%%%%%%%%%%%%%%%%%%%%%%%%%%
\subsection{Absence of the {\it visible sector}}
The next case will be concerned with studies of the only {\it hidden sector} fields on the thermoelectric properties. We assume
that $a^{(0)}_t = 0, ~E_x =0$. In this case one gets
\be
J^{x (0)}_{(F)} = \frac{1}{A_{hid}} \Big[
\frac{\alpha}{2} B_x  ~ \int dx ~Y_{hid} 
+ 4 \pi T~\xi_x \int dx \sqrt{\frac{\ga(x)}{\la(x)}}~\frac{\alpha}{2} b^{(0)}_t \Big],
\ee
and respectively for $J^{x (0)}_B$
\be
J^{x (0)}_{(B)} = \frac{1}{A_{hid}} \Big[
B_x ~\int dx ~Y_{hid} + 4 \pi T~\xi_x \int dx \sqrt{\frac{\ga(x)}{\la(x)}}~ b^{(0)}_t  \Big],
\ee
with the other quantities defined as
\be
v_0(hid) = \frac{1}{A_{hid}} \Big(
B_x \int dx \sqrt{\frac{\ga(x)}{\la(x)}} ~b^{(0)}_t + 4 \pi T~\xi_x ~\int dx \sqrt{\frac{\ga(x)}{\la(x)}} \Big),
\ee
\be
A_{hid} = \int dx \sqrt{\frac{\ga(x)}{\la(x)}} ~\int dx ~Y_{hid} - \Big( \int dx \sqrt{\frac{\ga(x)}{\la(x)}}~b^{(0)}_t \Big)^2,
\ee
\be
Y_{hid} = \frac{(\p_x \phi)^2}{\sqrt{\la(x) \ga(x)}} + b^{(0)2}_t + \frac{1}{4 { \la (x)}^{\frac{5}{2}} {\ga(x)}^{\frac{1}{2}}}
{( \p_x \la (x))}^2~
 v_0(hid).
\ee
In the considered case of {\it hidden sector} fields the DC coefficients reduce to the forms
\ben \label{hid}
\sigma_{FF} &=& 0,\\
\sigma_{FB} &=&  \frac{1}{A_{hid}} ~ \frac{\alpha}{2}
\int dx~ Y_{hid},  \\
\sigma_{BF} &=&  0, \\
\sigma_{BB} &=&  \frac{1}{A_{hid}}
\int dx ~Y_{hid},\\
\alpha_{(F)} &=& \frac{4 \pi}{A_{hid}} \int dx \sqrt{\frac{\ga(x)}{\la(x)}}  \frac{\alpha}{2} b^{(0)}_t,\\
\alpha_{(B)} &=& \frac{4 \pi}{A_{hid}} \int dx \sqrt{\frac{\ga(x)}{\la(x)}}
~ b^{(0)}_t, \\ 
\kappa &=& \frac{(4 \pi)^2 T}{A_{hid}} \int dx \sqrt{\frac{\ga(x)}{\la(x)}}.
\een
It is interesting to notice that despite of the absence of the {\it visible sector} fields, we get non-zero values of $\sigma_{FB} $ and $\alpha_{(F)}$, all of them proportional
to the $\alpha$-coupling constant. This situation has its roots in the definition of the gauge currents, in which we have both {\it visible} and {\it hidden} sector fields.

%%%%%%%%%%%%%%%%%%%%%%%%%%%%%%%%%%%%%%%%%%%%%%%%%%%%%%%%
\subsubsection{Non-interacting gauge fields}

In this case we neglect the interaction between two $U(1)$-gauge fields. Namely we put $\alpha$-coupling constant equal to zero. For the studied case one has
\ben \label{jfnon}
J^{x (0)}_{(F)} &=& \frac{1}{A_{\alpha =0}} \Bigg[
E_x  \Big[ \int dx ~Y_{\alpha =0} - \int dx \sqrt{\frac{\ga(x)}{\la(x)}}~b^{(0) 2}_t \Big] \\ \nonumber
&+& 4 \pi T~\xi_x \int dx \sqrt{\frac{\ga(x)}{\la(x)}}~a^{(0)}_t 
+ B_x ~ \int dx \sqrt{\frac{\ga(x)}{\la(x)}}~b^{(0)}_t ~ a^{(0)}_t  \Bigg],
\een
and respectively for $J^{x (0)}_B$
\ben \label{jbnon}
J^{x (0)}_{(B)} &=& \frac{1}{A_{\alpha =0}} \Bigg[
B_x ~\Big[ \int dx ~Y_{\alpha =0} - \int dx \sqrt{\frac{\ga(x)}{\la(x)}}~a^{(0) 2}_t  \Big] \\ \nonumber
&+& 4 \pi T~\xi_x \int dx \sqrt{\frac{\ga(x)}{\la(x)}}~ b^{(0)}_t
+ E_x ~\int dx \sqrt{\frac{\ga(x)}{\la(x)}}~a^{(0)}_t b^{(0)}_t  \Bigg].
\een
where we denoted
\ben \nonumber \label{vonon}
v_0 (\alpha =0)&=& \frac{1}{A_{\alpha =0}} \Big[ 
B_x ~\int dx \sqrt{\frac{\ga(x)}{\la(x)}} ~b^{(0)}_t + E_x~ \int dx \sqrt{\frac{\ga(x)}{\la(x)}} ~a^{(0)}_t \\
&+& 4 \pi T~\xi_x ~\int dx \sqrt{\frac{\ga(x)}{\la(x)}} \Big],
\een
\be
A_{\alpha =0} = \int dx \sqrt{\frac{\ga(x)}{\la(x)}} ~\int dx ~Y_{\alpha =0} - \Big( \int dx \sqrt{\frac{\ga(x)}{\la(x)}}~a^{(0)}_t \Big)^2  - \Big( \int dx \sqrt{\frac{\ga(x)}{\la(x)}}~b^{(0)}_t \Big)^2,
\ee 
\be
Y_{\alpha =0} = \frac{(\p_x \phi)^2}{\sqrt{\la(x) \ga(x)}} +  a^{(0)2}_t + b^{(0)2}_t + \frac{1}{4 { \la (x)}^{\frac{5}{2}} {\ga(x)}^{\frac{1}{2}}}
{( \p_x \la (x))}^2~
 v_0(\alpha=0).
\ee
The thermoelectric coefficients satisfy the relations
\ben 
\sigma_{FF} &=& \frac{1}{A_{\alpha =0}} \Big[
\int dx~ Y_{\alpha =0} - \int dx \sqrt{\frac{\ga(x)}{\la(x)}}~b^{(0) 2}_t \Big], \\
\sigma_{FB} &=&  \sigma_{BF}= \frac{1}{A_{\alpha =0}} 
~\int dx \sqrt{\frac{\ga(x)}{\la(x)}}~a^{(0)}_t~b^{(0) }_t, \\
\sigma_{BB} &=&  \frac{1}{A_{\alpha =0}} \Big[
\int dx ~Y - \int dx \sqrt{\frac{\ga(x)}{\la(x)}}~a^{(0) 2}_t \Big], \\
\alpha_{(F)} &=& \frac{4 \pi}{A_{\alpha =0}} \int dx \sqrt{\frac{\ga(x)}{\la(x)}}
~ a^{(0)}_t,\\
\alpha_{(B)} &=& \frac{4 \pi}{A_{\alpha =0}} \int dx \sqrt{\frac{\ga(x)}{\la(x)}}
~ b^{(0)}_t,\\
\kappa &=& \frac{(4 \pi)^2 T}{A_{\alpha =0}} \int dx \sqrt{\frac{\ga(x)}{\la(x)}}.
\een
In this case we achieve the full matrix of DC coefficients, with the equality $\sigma_{FB} =  \sigma_{BF}$.

As a general remark one finds that the form of $\kappa$ is the same in all cases (of course up to the concrete meaning of $A,~Y,~v_0$).

%%%%%%%%%%%%%%%%%%%%%%%%%%%%%%%%%%%%%%%%%%%%%%%%%%%%%%%%%%%%%%%%%%
%%%%%%%%%%%%%%%%%%%%%%%%%%%%%%%%%%%%%%%%%%%%%%%%%%%%%%%%%%%%%%
\section{Summary and conclusions}
\label{sum-concl}
In our paper we have studied the DC thermoelectric conductivities for the holographic model of Dirac semimetals. Our attitude to the problem was to implement the
top-down procedure, i.e., we started from fully quantum, consistent theory like string/M-theory, which ensured us that the predictions revealing from the holographic
correspondence were physical.

The gravity theory in AdS spacetime includes two $U(1)$-gauge fields. One of them is the ordinary Maxwell field, while the other is connected with
the {\it hidden sector}. Both fields interact with themselves via {\it kinetic mixing } term. Our main aim is to have an insight into the role of the auxiliary gauge 
field and $\alpha$-coupling constant on the physics. 

We search for the black brane response to the electric fields connected with the two $U(1)$-gauge fields and temperature gradient.
The foliation by hypersurfaces of constant radial coordinate enables us to derive the exact form of the Hamiltonian and equations of motion in
the considered phase space. On the other hand, an inspection of the Hamiltonian constraints authorises, to the leading order expansion of the linearised perturbations 
at the black brane event horizon, the derivation of Stokes equations for incompressible doubly charged fluid. Solving the aforementioned equations, one arrives at
the DC conductivities for the holographic Dirac semimetals.

We also address the question how the considered formalism of Stokes equations reconstructs the previously obtained results in the holographic
two current models \cite{seo17,rog18, rog18a}. Supposing that the {\it visible sector} charge is proportional to the {\it hidden sector} one, we derive the relation describing electric conductivity. It is in a very good agreement with the preceding ones.

As an example of the derived formalism we studied the case of one-dimensional periodic lattice, taking into account {\it visible}, {\it hidden} sector fields and the non-interacting 
gauge fields. In the first case we arrive at the relations derived in \cite{ban15}, confirming the right limit of our model. On the other hand, the inspection of the
{\it hidden sector} fields reveals that some kinetic coefficients bounded with the Maxwell field survives, i.e., $\sigma_{FB},~\alpha_{(F)}$ are not equal to zero. One can explain
this fact having in mind the adequate definitions, where the mixture of {\it visible } and {\it hidden} sector fields takes place. The outlasted thermoelectric conductivities
are proportional to $\alpha$-coupling constant. In the case of non-interacting gauge fields we obtained the full matrix of the conductivities. It is interesting to notice that
$\kappa$-coefficient, in all the elaborated examples, has the same form (up to the concrete meaning of $A,~Y,~v_0$).

%%%%%%%%%%%%%%%%%%%%%%%%%%%%%%%%%%%%%%%%
As far as the new directions for the further explorations is concerned, the non-linear time-dependent Navier-Stokes equations at the black object event horizon,
seems to be a natural generalisation 
of the researches. They should in principle give information for dual CFT, connected with AC thermoelectric conductivities and the influence of the {\it hidden sector}
on the time-dependent phenomena. On the other hand, inclusion of magnetic fields, both ordinary and connected with auxiliary gauge field, should have its
imprint in the physics. We hope to pay attention to the problems in question elsewhere.

%%%%%%%%%%%%%%%%%%%%%%%%%%%%%%%%%%%%%%%%%%%%%%%%%%%%%%%%%%%%%%%%%%
\acknowledgments
We thank K.I. Wysokinski for discussions on various occasions.

%%%%%%%%%%%%%%%%%%%%%%%%%%%%%%%%%%%%%%%%%%%%%%%%%%%%%%%%%%%%%%%%%%%%%%%%%%%%%%
%\paragraph{Note added.} This is also a good position for notes added
%after the paper has been written.


%%%%%%%%%%%%%%%%%%%%%%%%%%%%%%%%%%%%%%%%%%%%%%%%%%%%%%%%%%%%%%%%%%%%%%%%%%%%%%%
\end{document}